\documentclass[11pt,dvips]{article}
\usepackage{eurosym}
\usepackage{graphicx}
\usepackage{amsmath}
\usepackage{amssymb}
\usepackage{latexsym}
\usepackage{color}
\usepackage{epstopdf}

\input{tcilatex}

\begin{document}

\thispagestyle{empty}

\begin{center}
\vspace{1.8cm}

{\Large \textbf{Controlling Stationary One-Way Steering via Thermal Effects
in Optomechanics}} 

\vspace{1.5cm}

\textbf{Jamal El Qars}$^{a}${\footnote{%
email: \textsf{j.elqars@gmail.com}}}, \textbf{Mohammed Daoud}$^{b,c}${%
\footnote{%
email: \textsf{m$_{-}$daoud@hotmail.com}}} and \textbf{Rachid Ahl Laamara}$%
^{a,d} ${\footnote{%
email: \textsf{ahllaamara@gmail.com}}}

\vspace{0.5cm}

$^{a}$\textit{LPHE-MS, Faculty of Sciences, Mohammed V University, Rabat,
Morocco}\\[0.5em]
$^{b}$\textit{Abdus Salam International Centre for Theoretical Physics,
Miramare, Trieste, Italy}\\[0.5em]
$^{c}$\textit{Department of Physics, Faculty of Sciences, University Hassan
II, Casablanca, Morocco}\\[0.5em]
$^{d}$\textit{Centre of Physics and Mathematics (CPM), Mohammed V
University, Rabat, Morocco}\\[0.5em]

\vspace{3cm} \textbf{Abstract}
\end{center}

Quantum steering is a kind of quantum correlations stronger than
entanglement but weaker than Bell-nonlocality. In an optomechanical system
pumped by squeezed light and driven in the red sideband, we study-under
thermal effects-stationary Gaussian steering and its asymmetry of two
mechanical modes. In the resolved sideband regime using experimentally
feasible parameters, we show that Gaussian steering can be created by
quantum fluctuations transfer from the squeezed light to the two mechanical
modes. Moreover, one-way steering can be observed by controlling the
squeezing degree or the environmental temperature. A comparative study
between Gaussian steering and Gaussian R\'{e}nyi-2 entanglement of the two
considered modes shows on one hand that both steering and entanglement
suffer from a \textit{sudden death}-like phenomenon with early vanishing of
steering in various circumstances. On the other hand, steering is found
stronger than entanglement, however, remains constantly upper bounded by
Gaussian R\'{e}nyi-2 entanglement, and decays rapidly to zero under thermal
noise. 
\newpage

\section{Introduction}

In their seminal 1935 paper \cite{EPR}, Einstein, Podolsky, and Rosen (EPR)
have highlighted that, when two spatially separated particles are entangled,
two arbitrary local measurements performed upon one particle induce
immediate nonlocal effects on the second, where this end may be left in
states with two different wave-functions. To capture the essence of this 
\textit{spooky action-at-a-distance} phenomenon on one hand, and to
generalize the EPR paradox on the other hand, Schr\"{o}dinger has originally
introduced the concept of \textit{steer} as an exotic quantum effect
allowing remotely preparation of quantum states by local operations \cite%
{Schrodinger}. Recently, quantum steering or EPR steering was defined as a
kind of \textit{non-separable} quantum correlations \cite{WJD}, weaker than
Bell nonlocality \cite{Bell}, but stronger than entanglement \cite{The
Horodeckis}. Precisely, for a given quantum bipartite state, the violation
of Bell inequality implies EPR steering in both directions, while, steering
at least in one direction implies that the state is entangled \cite{WJD}.
Interestingly enough, moving from entanglement to EPR steering to
Bell-nonlocality; requires decreasing the number of observers and
apparatuses that must be trusted \cite{WJD}. In contrast, the corresponding
protocols were shown to be progressively less robust against thermal noise
for projective measurements \cite{Saunders}.

From an operational perspective in quantum information theory, EPR steering
corresponds to an entanglement verification task, i.e. it certifies the
existence of entanglement between two parties, assuming trusted measurements
only on one side. Specifically, if Alice and Bob jointly share a steerable
state at least in one direction (say from Alice to Bob), then, Alice can
convince Bob who does not trust her that their shared state is entangled, by
performing local operations and classical communication.

On the basis of the uncertainty principle, a quantitative criterion to test
the experimental EPR paradox was proposed in \cite{Reid1}. Further, it has
been proven that the violation of such criterion under Gaussian measurements
witnesses EPR steering \cite{WJD}, where the first verification was realised
in \cite{Ou}.

To detect quantum steering, several inequalities are available \cite%
{Steering inq}, where their violation certifies EPR steering \cite{Brunner},
however, they cannot quantify it \cite{Kogias2}. In this sense, two steering
quantifiers were proposed, i.e. steering weight \cite{Skrzypczyk}, and the
steering robustness \cite{Watrous}. Unfortunately, both quantifiers are not
evaluated in a compact form, where they can only be computed numerically by
semidefinite programming \cite{Kogias2}, while, for two-mode Gaussian
states, a computable measure was developed \cite{Kogias1}.

Actually, our knowledge concerning the generation, the detection and also
the quantification of EPR steering has significantly advanced, where intense
efforts have been made in the last decade to study such phenomenon
theoretically \cite{Olsen1,G1} as well as experimentally \cite%
{G2,Evans,Bowles}.

Importantly, the distinctive feature of EPR steering-unlike entanglement and
Bell nonlocality-that is intrinsically asymmetric, i.e. an entangled quantum
state may be steerable from Alice to Bob, but not vice versa, which is
commonly known as one-way steering \cite{WJD,Kogias1}. Apart from its
fundamental relevance, one-way steering has attracted a great deal of
attention \cite{Bowles,oneway,Wollmann}, where it has been appreciated as a
resource for various quantum information protocols, e.g. one-sided
device-independent quantum cryptography \cite{1sDI}, secure quantum
teleportation \cite{Reid's}, and subchannel discrimination \cite{Watrous}.

Restricting to Gaussian states and measurements, experimental one-way
steering was carried out in \cite{Handchen} and later in \cite{Armstrong},
which answer genuinely the question already raised by Wiseman \textit{et al}%
, i.e. are there entangled states which are one-way steerable \cite{WJD}. We
recall that pure entangled states cannot exhibit one-way steering, where
they can always be transformed to a symmetric form via local basis change
using the Schmidt decomposition \cite{Bowles}.

We note here that in an optical setup, one-way steering has been proven
theoretically and then experimentally observed under Gaussian and
non-Gaussian measurements in Ref. \cite{Wollmann}. While, in hybrid
optomechanical systems, such behavior has been studied considering Gaussian
measurements on Gaussian states in Refs. \cite{QSOS,Huatang}. In this paper,
using two spatially separated optomechanical cavities coupled to two-mode
squeezed light and driven in the red sideband, we investigate-under thermal
effects-Gaussian EPR steering of two non-interacting mechanical modes, where
a particular attention is dedicated to the one-way steering behavior. Also,
we compare the Gaussian steering of the two considered modes with their
Gaussian R\'{e}nyi-2 entanglement \cite{AGS}.

Over the past decades, substantial efforts in optomechanics have been made
to test quantum effects \cite{Meystre,Aspelmeyer}. This includes, cooling of
a mechanical oscillator to its ground state \cite{Cooling}, optomechanically
induced transparency \cite{Agarwal}, quantum squeezing \cite{Liao},
macroscopic superposition state \cite{Marshall}, backaction-evading
measurement \cite{Korppi}, quantum entanglement \cite{entanglement} and also
quantum steering \cite{QSOS,Huatang}.

This paper is organized as follows. In Sec. \ref{sec2}, we introduce the
basic optomechanical system involving two optical modes and two mechanical
modes. Next, using the dynamics based on the quantum Langevin equations, we
derive in the resolved sideband regime the covariance matrix describing
stationary four-mode Gaussian states of the system. In Sec. \ref{sec3},
focusing on the mechanical modes, we quantify their Gaussian steering as
well as their Gaussian R\'{e}nyi-2 entanglement. We study and compare these
two different aspects of non-separable quantum correlations under thermal
noises induced by the squeezing effect and the environmental temperatures.
In Sec. \ref{sec4}, we draw our conclusions.

\section{A Fabry-P\'{e}rot double-cavity optomechanical system \label{sec2}}

\subsection{Model and Hamiltonian}

\begin{figure}[tbh]
\centerline{\includegraphics[width=13cm]{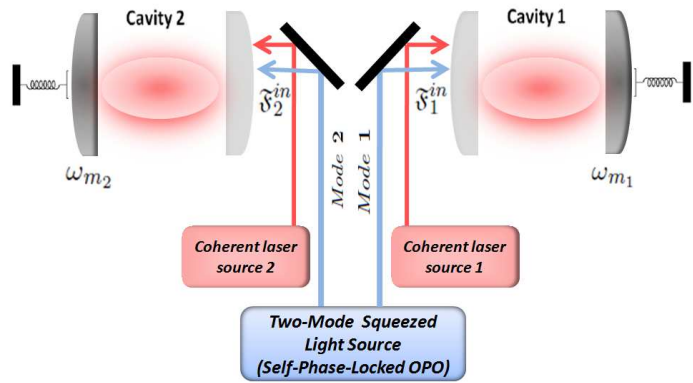}}
\caption{A double-cavity optomechanical system driven in the red sideband
and fed by two-mode squeezed light with input noise operators $\mathfrak{F}%
_{1,2}^{in}$. Each movable mirror oscillates at frequency $\protect\omega %
_{m_{j}}$.}
\label{Fig.1}
\end{figure}
\noindent The proposed optomechanical system comprises two Fabry-P\'{e}rot
cavities (Fig. \ref{Fig.1}), where each cavity is composed by two mirrors.
The first one is fixed and partially transmitting, while, the second is
movable and perfectly reflecting. The $j\mathrm{th}$ cavity has length $%
l_{j} $ and driven by coherent laser with input power ${\wp }_{j}$, phase $%
\varphi _{j}$ and frequency $\omega _{L_{j}}$. Also, the two cavities are
fed by two-mode squeezed light of frequency $\omega _{s}$. The first(second)
squeezed mode is sent towards the first(second) cavity. \newline
The two Fabry-P\'{e}rot cavities can be described by the following
Hamiltonian ($\hbar =1$) \cite{Law} 
\begin{equation}
\mathcal{\hat{H}=}\sum_{j=1}^{2}\Big(\mathcal{\hat{H}}_{c_{j}}\mathcal{+\hat{%
H}}_{m_{j}}\mathcal{+\hat{H}}_{coup_{j}}\mathcal{+\hat{H}}_{drive_{j}}\Big),
\label{E1}
\end{equation}%
where $\mathcal{\hat{H}}_{c_{j}}=\omega _{c_{j}}a_{j}^{\dag }a_{j}$ is the
free Hamiltonian of the $j\mathrm{th}$ cavity mode with annihilation
operator $a_{j}$, frequency $\omega _{c_{j}}$ and decay rate $\kappa
_{c_{j}} $, while, $\mathcal{\hat{H}}_{m_{j}}=\omega _{m_{j}}b_{j}^{\dag
}b_{j}$ is the free Hamiltonian of the $j\mathrm{th}$ movable mirror modeled
as a single-mode quantum harmonic oscillator with annihilation operator $%
b_{j}$, an effective mass $\mu _{j}$, frequency $\omega _{m_{j}}$ and
damping rate $\gamma _{m_{j}}$. The term $\mathcal{\hat{H}}_{coup_{j}}=\chi
_{0_{j}}a_{j}^{\dag }a_{j}(b_{j}^{\dag }+b_{j})$ is the radiation pressure
coupling between the $j\mathrm{th}$ cavity mode and its corresponding
mechanical mode, with coupling rate $\chi _{0_{j}}=\left( \omega
_{c_{j}}/l_{j}\right) \sqrt{\hbar /\mu _{_{j}}\omega _{m_{j}}}$. Finally, $%
\mathcal{\hat{H}}_{drive_{j}}=\varepsilon _{j}(a_{j}^{\dag }e^{i\varphi
_{j}}e^{-i\omega _{L_{j}}}+a_{j}e^{-i\varphi _{j}}e^{i\omega _{L_{j}}})$ is
the coupling between the $j\mathrm{th}$ laser and the $j\mathrm{th}$ cavity,
with coupling strength $\varepsilon _{j}=\sqrt{2\kappa _{c_{j}}{\wp }%
_{j}/\hbar \omega _{L_{j}}}$. 

\subsection{System dynamics}

The system at hand is a dissipative-noisy optomechanical system. So, the
dynamics can be conveniently described by the quantum Langevin equation,i.e.$%
\partial _{t}\mathcal{O}=\frac{1}{i\hbar }\left[ \mathcal{O},\mathcal{\hat{H}%
}\right] +dissipation$ $and$ $noise$ $terms$ ($\mathcal{O\equiv }a_{j},b_{j}$%
) \cite{Paternostro}. Thus, in a frame rotating with $\omega _{L_{j}}$, we
obtain the following nonlinear equations 
\begin{eqnarray}
\partial _{t}b_{j} &=&-\left( \frac{\gamma _{m_{j}}}{2}+i\omega
_{m_{j}}\right) b_{j}-i\chi _{0_{j}}a_{j}^{\dag }a_{j}+\sqrt{\gamma _{m_{j}}}%
\zeta _{j}^{in},  \label{E2} \\
\partial _{t}a_{j} &=&-\left( \frac{\kappa _{c_{j}}}{2}-i\Delta _{j}\right)
a_{j}-i\chi _{0_{j}}a_{j}(b_{j}^{\dag }+b_{j})-i\varepsilon _{j}e^{i\varphi
_{j}}+\sqrt{\kappa _{c_{j}}}\mathfrak{F}_{j}^{in},  \label{E3}
\end{eqnarray}%
with $\Delta _{j}=\omega _{L_{j}}-\omega _{c_{j}}$ is the $j\mathrm{th}$
laser detuning \cite{Aspelmeyer}. $\zeta _{j}^{in}$ is the zero-mean
Brownian noise operator affecting the $j\mathrm{th}$ movable mirror. It is
not in general $\delta $-correlated, exhibiting a non-Markovian correlation
function between two instants $t$ and $t^{\prime }$ \cite{Genes}. However,
oscillators with large mechanical quality factor $\mathcal{Q}_{m_{j}}=\omega
_{m_{j}}/\gamma _{m_{j}}\gg 1$ allows recovering the Markovian process and,
then, quantum effects can be reached. In this limit, we have the following
nonzero time-domain correlation functions \cite{Benguria} 
\begin{equation}
\langle \zeta _{j}^{in\dag }(t)\zeta _{j}^{in}(t^{\prime });\zeta
_{j}^{in}(t)\zeta _{j}^{in\dag }(t^{\prime })\rangle =\Big(%
n_{th,j};n_{th,j}+1\Big)\delta (t-t^{\prime }),\text{ \ }j=1,2,  \label{E5}
\end{equation}%
with $n_{th,j}=(e^{\hbar \omega _{m_{j}}/k_{B}T_{j}}-1)^{-1}$ is the $j%
\mathrm{th}$ mean number of phonons. $T_{j}$ and $k_{B}$ are the temperature
of the $j\mathrm{th}$ mirror environment and the Boltzmann constant. In Eq. (%
\ref{E3}), $\mathfrak{F}_{j}^{in}$ is the $j\mathrm{th}$ zero mean input
squeezed noise operator, with the following nonzero time-domain correlation
functions \cite{Paternostro} 
\begin{eqnarray}
\langle \delta \mathfrak{F}_{j}^{in^{\dag }}(t)\delta \mathfrak{F}%
_{j}^{in}(t^{\prime });\delta \mathfrak{F}_{j}^{in}(t)\delta \mathfrak{F}%
_{j}^{in^{\dag }}(t^{\prime })\rangle &=&\Big(N;N+1\Big)\delta (t-t^{\prime
}),\text{ \ }j=1,2,  \label{E6} \\
\langle \delta \mathfrak{F}_{j}^{in}(t)\delta \mathfrak{F}_{j^{\prime
}}^{in}(t^{\prime });\delta \mathfrak{F}_{j}^{in^{\dag }}(t)\delta \mathfrak{%
F}_{j^{\prime }}^{in^{\dag }}(t^{\prime })\rangle &=&\Big(Me^{-i\omega
_{m}\left( t+t^{\prime }\right) };Me^{i\omega _{m}\left( t+t^{\prime
}\right) }\Big)\delta (t-t^{\prime }),\text{ \ }j\neq j^{\prime }=1,2,
\label{E7}
\end{eqnarray}%
with $N=\mathrm{sinh}^{\mathrm{2}}r$, $M=\mathrm{sinh}r\mathrm{cosh}r$, $r$
is the squeezing parameter (we have assumed that $\omega _{m_{1,2}}=\omega
_{m}$). We note that optimal transfer from the squeezed light to the
mechanical degrees of freedom can be achieved when the frequency of the
squeezing is resonant with those of the cavities,i.e. $\omega _{s}=\omega
_{c_{j}}$ \cite{Parkin and Kimble}. 

\subsection{Linearization of the dynamics around the steady-states}

The equations (\ref{E2})-(\ref{E3}) are nonlinear due to the quadratic terms 
$a_{j}^{\dag }a_{j}$, $a_{j}b_{j}^{\dag }$ and $a_{j}b_{j}$, then, they
cannot be solved exactly \cite{Genes}. Assuming weak coupling between the $j%
\mathrm{th}$ cavity mode and its associated mechanical mode, the
fluctuations $\delta a_{j}$, $\delta b_{j}$ are much smaller than the
steady-state mean values $\langle a_{j}\rangle $ and $\langle b_{j}\rangle $%
. So, we can linearize the dynamics around the steady state, where each
operator can be written as sum of its steady-state mean value and a small
fluctuation with zero mean value,i.e. $\mathcal{O}_{j}=\langle \mathcal{O}%
_{j}\rangle +\delta \mathcal{O}_{j}$ ($\mathcal{O}_{j}\equiv a_{j},b_{j}$) 
\cite{Genes}. The mean values are obtained by setting the time derivatives
to zero and factorizing the averages in Eqs. (\ref{E2})-(\ref{E3}). Thus, we
get $\langle a_{j}\rangle \equiv a_{js}=\frac{-2i\varepsilon _{j}e^{i\varphi
_{j}}}{\kappa _{c_{j}}-2i\Delta _{j}^{\prime }}$ and $\langle b_{j}\rangle
\equiv b_{js}=\frac{-2i\chi _{0_{j}}\left\vert a_{js}\right\vert ^{2}}{%
\gamma _{m_{j}}+2i\omega _{m_{j}}}$ with $\Delta _{j}^{\prime }$ $=$ $\Delta
_{j}$ $-\chi _{0_{j}}(b_{js}{}^{\ast }+b_{js})$ is the $j\mathrm{th}$
effective detuning \cite{Aspelmeyer}. To simplify further our purpose, we
assume that the two cavities are intensely driven,i.e. $\left\vert
a_{js}\right\vert \gg 1$ by lasers with large powers $\wp _{1,2}$ \cite%
{Genes}. So, the terms $\delta a_{j}^{\dag }\delta a_{j}$, $\delta
a_{j}\delta b_{j}$ and $\delta a_{j}\delta b_{j}^{\dag }$ can be safely
neglected, leading to 
\begin{eqnarray}
\delta \dot{b}_{j} &=&-\left( \frac{\gamma _{m_{j}}}{2}+i\omega
_{m_{j}}\right) \delta b_{j}+\chi _{j}\left( \delta a_{j}-\delta a_{j}^{\dag
}\right) +\sqrt{\gamma _{m_{j}}}\zeta _{j}^{in},  \label{E11} \\
\delta \dot{a}_{j} &=&-\left( \frac{\kappa _{c_{j}}}{2}-i\Delta _{j}^{\prime
}\right) \delta a_{j}-\chi _{j}\left( \delta b_{j}+\delta b_{j}^{\dag
}\right) +\sqrt{\kappa _{c_{j}}}\delta \mathfrak{F}_{j}^{in},  \label{E12}
\end{eqnarray}%
$\chi _{j}$ $=\chi _{0_{j}}\left\vert a_{js}\right\vert $ being the $j%
\mathrm{th}$ effective coupling \cite{Aspelmeyer}. Notice that Eqs. (\ref%
{E11})-(\ref{E12}) are obtained by setting $a_{js}=-i\left\vert
a_{js}\right\vert $ or equivalently to $\tan \varphi _{j}=-2\Delta
_{j}^{\prime }/\kappa _{c_{j}}$. Next, we introduce the operators $\delta 
\tilde{b}_{j}=\delta b_{j}e^{i\omega _{m_{j}}t}$ and $\delta \tilde{a}%
_{j}=\delta a_{j}e^{-i\Delta _{j}^{\prime }t}$ and we assume that the system
is driven in \textit{the red sideband} ($\Delta _{j}^{\prime }=-\omega
_{m_{j}}$), which is appropriate for quantum-state transfer \cite{Aspelmeyer}%
. Moreover, in the resolved-sideband regime,i.e. $\omega _{m_{j}}\gg \kappa
_{c_{j}}$, the rotating wave approximation (RWA) allows to drop terms
rotating at $\pm 2\omega _{m_{j}}$ \cite{RWA}. Thus, one has 
\begin{eqnarray}
\delta \dot{\tilde{b}}_{j} &=&-\frac{\gamma _{m_{j}}}{2}\delta \tilde{b}%
_{j}+\chi _{j}\delta \tilde{a}_{j}+\sqrt{\gamma _{m_{j}}}\tilde{\zeta}%
_{j}^{in},  \label{E15} \\
\delta \dot{\tilde{a}}_{j} &=&-\frac{\kappa _{c_{j}}}{2}\delta \tilde{a}%
_{j}-\chi _{j}\delta \tilde{b}_{j}+\sqrt{\kappa _{c_{j}}}\delta \mathfrak{%
\tilde{F}}_{j}^{in}.  \label{E16}
\end{eqnarray}

\subsection{Four-mode covariance matrix}

Using Eqs. (\ref{E15})-(\ref{E16}) and the quadratures position and momentum
of the $j\mathrm{th}$ mechanical(optical) mode $\delta \tilde{q}%
_{m_{j}}=(\delta \tilde{b}_{j}^{\dag }+\delta \tilde{b}_{j})/\sqrt{2}$ and $%
\delta \tilde{p}_{m_{j}}=i(\delta \tilde{b}_{j}^{\dag }-\delta \tilde{b}%
_{j})/\sqrt{2}$ $\Big(\delta \tilde{q}_{c_{j}}=(\delta \tilde{a}_{j}^{\dag
}+\delta \tilde{a}_{j})/\sqrt{2}$ and $\delta \tilde{p}_{c_{j}}=i(\delta 
\tilde{a}_{j}^{\dag }-\delta \tilde{a}_{j})/\sqrt{2}\Big)$ with their
corresponding input mechanical(optical) noise operators $\delta \tilde{q}%
_{m_{j}}^{in}=(\tilde{\zeta}_{j}^{in\dagger }+\tilde{\zeta}_{j}^{in})/\sqrt{2%
}$ and $\delta \tilde{p}_{m_{j}}^{in}=i(\tilde{\zeta}_{j}^{in\dagger }-%
\tilde{\zeta}_{j}^{in})/\sqrt{2}$ $\Big(\delta \tilde{q}_{c_{j}}^{in}=(%
\delta \mathfrak{\tilde{F}}_{j}^{in\dag }+\delta \mathfrak{\tilde{F}}%
_{j}^{in})/\sqrt{2}$ and $\delta \tilde{p}_{c_{j}}^{in}=i(\delta \mathfrak{%
\tilde{F}}_{j}^{in\dag }-\delta \mathfrak{\tilde{F}}_{j}^{in})/\sqrt{2}\Big)$%
, we obtain 
\begin{eqnarray}
\partial _{t}\delta \tilde{q}_{m_{j}} &=&-\frac{\gamma _{m_{j}}}{2}\delta 
\tilde{q}_{m_{j}}+\chi _{j}\delta \tilde{q}_{c_{j}}+\sqrt{\gamma _{m_{j}}}%
\delta \tilde{q}_{m_{j}}^{in},  \label{E17} \\
\partial _{t}\delta \tilde{p}_{m_{j}} &=&-\frac{\gamma _{m_{j}}}{2}\delta 
\tilde{p}_{m_{j}}+\chi _{j}\delta \tilde{p}_{c_{j}}+\sqrt{\gamma _{m_{j}}}%
\delta \tilde{p}_{m_{j}}^{in},  \label{E18} \\
\partial _{t}\delta \tilde{q}_{c_{j}} &=&-\frac{\kappa _{c_{j}}}{2}\delta 
\tilde{q}_{c_{j}}-\chi _{j}\delta \tilde{q}_{m_{j}}+\sqrt{\kappa _{c_{j}}}%
\delta \tilde{q}_{c_{j}}^{in},  \label{E19} \\
\partial _{t}\delta \tilde{p}_{c_{j}} &=&-\frac{\kappa _{c_{j}}}{2}\delta 
\tilde{p}_{c_{j}}-\chi _{j}\delta \tilde{p}_{m_{j}}+\sqrt{\kappa _{c_{j}}}%
\delta \tilde{p}_{c_{j}}^{in},  \label{E20}
\end{eqnarray}
which can be written as $\partial _{t}\tilde{u}=\mathcal{A}\tilde{u}+\tilde{n%
}$, with $\tilde{u}^{\mathrm{T}}=(\delta \tilde{q}_{m_{1}},\delta \tilde{p}%
_{m_{1}},\delta \tilde{q}_{m_{2}},\delta \tilde{p}_{m_{2}},\delta \tilde{q}%
_{c_{1}},\delta \tilde{p}_{c_{1}},\delta \tilde{q}_{c_{2}},\delta \tilde{p}%
_{c_{2}})$, $\tilde{n}^{\mathrm{T}}=(\delta \tilde{q}_{m_{1}}^{in},\delta 
\tilde{p}_{m_{1}}^{in},\delta \tilde{q}_{m_{2}}^{in},\delta \tilde{p}%
_{m_{2}}^{in},\delta \tilde{q}_{c_{1}}^{in},\delta \tilde{p}%
_{c_{1}}^{in},\delta \tilde{q}_{c_{2}}^{in},\delta \tilde{p}_{c_{2}}^{in})$
and $\mathcal{A=}\left( 
\begin{array}{cc}
\mathcal{A}_{\gamma _{m}} & \mathcal{A}_{\chi _{+}} \\ 
\mathcal{A}_{\chi _{-}} & \mathcal{A}_{\kappa _{c}}%
\end{array}%
\right) $, wherein the $2\times 2$ blocks matrices $\mathcal{A}_{\gamma
_{m}} $, $\mathcal{A}_{\kappa _{c}}$ and $\mathcal{A}_{\chi _{\pm }}$ are
respectively given by $\mathcal{A}_{\gamma _{m}}=\mathrm{diag}(-\frac{\gamma
_{m_{1}}}{2},-\frac{\gamma _{m_{1}}}{2},-\frac{\gamma _{m_{2}}}{2},-\frac{%
\gamma _{m_{2}}}{2})$, $\mathcal{A}_{\kappa _{c}}=\mathrm{diag}(-\frac{%
\kappa _{c_{1}}}{2},-\frac{\kappa _{c_{1}}}{2},-\frac{\kappa _{c_{2}}}{2},-%
\frac{\kappa _{c_{2}}}{2})$ and $\mathcal{A}_{\chi _{\pm }}=\mathrm{diag}%
(\pm \chi _{1},\pm \chi _{1},\pm \chi _{2},\pm \chi _{2})$.\newline
Since the dynamics is linearized and $\zeta _{j}^{in}$ and $\mathfrak{F}%
_{j}^{in}$ are zero-mean quantum Gaussian noises, \textit{the steady-state}
of the quantum fluctuations is a zero-mean quadripartite Gaussian state and
then, it can be described by its $8\times 8$ covariance matrix $\mathcal{V}$
defined as $\mathcal{V}_{kk^{\prime }}=\left( \langle \tilde{u}_{k}(\infty )%
\tilde{u}_{k^{\prime }}(\infty )+\tilde{u}_{k^{\prime }}(\infty )\tilde{u}%
_{k}(\infty )\rangle \right) /2$ \cite{Genes}. \newline
Using standard approaches \cite{Genes,Parks}, one can determine the matrix $%
\mathcal{V}$ by solving the Lyapunov equation 
\begin{equation}
\mathcal{AV+VA}^{\mathrm{T}}=-\mathcal{D},  \label{Lyapunov}
\end{equation}%
where $\mathcal{D}$ the diffusion matrix defined by $\mathcal{D}_{jj^{\prime
}}\delta (t-t^{\prime })=\left( \langle \tilde{n}_{j}(t)\tilde{n}_{j^{\prime
}}(t^{\prime })+\tilde{n}_{j^{\prime }}(t^{\prime })\tilde{n}_{j}(t)\rangle
\right)/2$ can be written as $\mathcal{D=D}_{\gamma _{m}}\oplus \mathcal{D}%
_{\kappa _{c}}$ with $\mathcal{D}_{\gamma _{m}}=\mathrm{diag}\Big(\gamma
_{m_{1}}(n_{th,1}+\frac{1}{2}),\gamma _{m_{1}}(n_{th,1}+\frac{1}{2}),\gamma
_{m_{2}}(n_{th,2}+\frac{1}{2}),\gamma _{m_{2}}(n_{th,2}+\frac{1}{2})\Big)$
and 
\begin{equation}
\mathcal{D}_{\kappa _{c}}=\left( 
\begin{array}{cccc}
\frac{\kappa _{c_{1}}}{2}\cosh (2r) & 0 & \frac{\sqrt{\kappa _{c_{1}}\kappa
_{c_{2}}}}{2}\sinh (2r) & 0 \\ 
0 & \frac{\kappa _{c_{1}}}{2}\cosh (2r) & 0 & \frac{-\sqrt{\kappa
_{c_{1}}\kappa _{c_{2}}}}{2}\sinh (2r) \\ 
\frac{\sqrt{\kappa _{c_{1}}\kappa _{c_{2}}}}{2}\sinh (2r) & 0 & \frac{\kappa
_{c_{2}}}{2}\cosh (2r) & 0 \\ 
0 & \frac{-\sqrt{\kappa _{c_{1}}\kappa _{c_{2}}}}{2}\sinh (2r) & 0 & \frac{%
\kappa _{c_{2}}}{2}\cosh (2r)%
\end{array}%
\right) .  \label{D-k}
\end{equation}%
The covariance matrix (CM) $\mathcal{V}$ solution of Eq. (\ref{Lyapunov}),
can be expressed as 
\begin{equation}
\mathcal{V=}\left[ \mathcal{V}_{ij}\right] _{8\times 8}\mathcal{=}\left( 
\begin{array}{cccc}
\mathcal{V}_{m_{1}} & \mathcal{V}_{m_{12}} & \mathcal{V}_{m_{1}c_{1}} & 
\mathcal{V}_{m_{1}c_{2}} \\ 
\mathcal{V}_{m_{12}}^{\mathrm{T}} & \mathcal{V}_{m_{2}} & \mathcal{V}%
_{m_{2}c_{1}} & \mathcal{V}_{m_{2}c_{2}} \\ 
\mathcal{V}_{m_{1}c_{1}}^{\mathrm{T}} & \mathcal{V}_{m_{2}c_{1}}^{\mathrm{T}}
& \mathcal{V}_{c_{1}} & \mathcal{V}_{c_{12}} \\ 
\mathcal{V}_{m_{1}c_{2}}^{\mathrm{T}} & \mathcal{V}_{m_{2}c_{2}}^{\mathrm{T}}
& \mathcal{V}_{c_{12}}^{\mathrm{T}} & \mathcal{V}_{c_{2}}%
\end{array}%
\right) ,  \label{CM}
\end{equation}%
where the $2\times 2$ blocks matrices $\mathcal{V}_{m_{1}}$ and $\mathcal{V}%
_{m_{2}}$ ($\mathcal{V}_{c_{1}}$ and $\mathcal{V}_{c_{2}}$) represent the
first and second mechanical(optical) modes, while their correlations are
described by $\mathcal{V}_{m_{12}}$($\mathcal{V}_{c_{12}}$). Moreover, the
matrix $\mathcal{V}_{m_{i}c_{j}}$ ($i,j\in \{1,2\}$) describes the
correlations between the $i\mathrm{th}$ mechanical mode and the $j\mathrm{th}
$ optical mode.\newline
Since we are interested in the Gaussian steering of the two mechanical modes
labeled as $A$ and $B$, their covariance matrix $\mathcal{V}_{m}$ can be
obtained by tracing over the uninteresting block matrices in (\ref{CM}).
Thus, one has 
\begin{equation}
\mathcal{V}_{m}=\left( 
\begin{array}{cc}
\mathcal{V}_{m_{1}} & \mathcal{V}_{m_{12}} \\ 
\mathcal{V}_{m_{12}}^{\mathrm{T}} & \mathcal{V}_{m_{2}}%
\end{array}%
\right) ,  \label{MCM}
\end{equation}%
with $\mathcal{V}_{m_{1}}=\mathrm{diag}(\upsilon _{1},\upsilon _{1})$, $%
\mathcal{V}_{m_{2}}=\mathrm{diag}(\upsilon _{2},\upsilon _{2})$ and $%
\mathcal{V}_{m_{12}}=\mathrm{diag}(\upsilon _{12},-\upsilon _{12})$. \newline
Assuming identical dampings,i.e.$\gamma _{m_{1,2}}=\gamma _{m}$ and $\kappa
_{c_{1,2}}=\kappa _{c}$, the matrix elements $\upsilon _{1}$, $\upsilon _{2}$
and $\upsilon _{12}$ are 
\begin{eqnarray}
\upsilon _{j} &=&\frac{\left( 1+2n_{th,j}\right) \left( 1+\tau +\tau 
\mathcal{C}_{j}\right) +\mathcal{C}_{j}\cosh (2r)}{2\left( 1+\tau \right)
\left( 1+\mathcal{C}_{j}\right) },\text{ for \ }j=1,2,  \label{nu2} \\
\upsilon _{12} &=&\frac{\sqrt{\mathcal{C}_{1}\mathcal{C}_{2}}\left( 1+\tau
\right) \sinh (2r)}{\left( 2+\mathcal{C}_{1}+\mathcal{C}_{2}\right) \left(
1+\tau \right) ^{2}+\frac{\tau }{2}\left( \mathcal{C}_{1}-\mathcal{C}%
_{2}\right) ^{2}},  \label{nu3}
\end{eqnarray}%
where $\tau =\gamma _{m}/\kappa _{c}$ is the damping ratio and $\mathcal{C}%
_{j}$ is the $j\mathrm{th}$ optomechanical cooperativity defined by \cite{OC}
\begin{equation}
\mathcal{C}_{j}=4\chi _{j}^{2}/\gamma _{m}\kappa _{c}=\frac{8\omega
_{c_{j}}^{2}\wp _{j}}{\gamma _{m}\mu _{_{j}}\omega _{m_{j}}\omega
_{L_{j}}l_{j}^{2}\left[ \left( \frac{\kappa _{c}}{2}\right) ^{2}+\omega
_{_{m_{j}}}^{2}\right] },  \label{OC}
\end{equation}%
Based on Eqs. (\ref{MCM})-(\ref{nu3}), we remark that if at least one of the
three parameters $\mathcal{C}_{1}$, $\mathcal{C}_{2}$ or $r$ is zero, then $%
\upsilon _{12}=0$, $\det \mathcal{V}_{m_{12}}$ will be as well. This implies
that the CM $\mathcal{V}_{m}$ (\ref{MCM}) describing the two mechanical
modes $A$ and $B$ is a Gaussian product states \cite{G(1)}. So, the two
modes $A$ and $B$ remain unentangled and, therefore, they can not be
steerable neither from $A\rightarrow B$ nor from $B\rightarrow A$ \cite%
{Kogias1}. This is because $\det \mathcal{V}_{m_{12}}$ $<0$ is a necessary
condition for a two-mode Gaussian states $\varrho _{AB}$ to be entangled 
\cite{AL}. Finally, it should be aware that the stability conditions have
been verified, where they are always satisfied in the chosen parameter
regime. This can be explained by the fact that the two cavities are driven
in the red sideband \cite{Genes}.

\section{Gaussian EPR steering vs Gaussian R\'{e}nyi-2 entanglement}

\label{sec3}

\subsection{Gaussian EPR steering}

To study the Gaussian EPR steering of the two mechanical modes $A$ and $B$,
we adopt the measure proposed in \cite{Kogias1}. For arbitrary \textit{%
two-mode Gaussian states} with covariance matrix $\mathcal{V}_{m}$ (\ref{MCM}%
), Alice can steer the Bob's states by performing Gaussian measurements, if
the following condition is violated \cite{WJD} 
\begin{equation}
\mathcal{V}_{m}+i(0_{A}\oplus \Omega _{B})\geqslant 0,  \label{Schro-Rober}
\end{equation}%
where $0_{A}$ is a $2\times 2$ null matrix and $\Omega _{B}$ $=\left( 
\begin{array}{cc}
0 & 1 \\ 
-1 & 0%
\end{array}%
\right) $ is the $B$-mode symplectic matrix \cite{Kogias1}. Henceforth, the
violation of the condition (\ref{Schro-Rober}) is necessary and sufficient
for Gaussian $A\rightarrow B$ steerability \cite{Kogias1}.\newline
To quantify how much an arbitrary bipartite Gaussian state with CM $\mathcal{%
V}_{m}$ is steerable under Gaussian measurements on Alice's side, Kogias 
\textit{et al} have been proposed the following measure \cite{Kogias1} 
\begin{equation}
\mathcal{G}^{A\rightarrow B}(\mathcal{V}_{m}):=\max \{0,-\ln (\eta ^{B})\},
\label{arb}
\end{equation}%
where $\eta ^{B}=\sqrt{\det (\mathcal{V}_{m_{2}}-\mathcal{V}_{m_{12}}^{%
\mathrm{T}}\mathcal{V}_{m_{1}}^{-1}\mathcal{V}_{m_{12}})}$. \newline
The steering $\mathcal{G}^{A\rightarrow B}$ is monotone under Gaussian LOCC 
\cite{Lami1}, it quantifies the amount by which the condition (\ref%
{Schro-Rober}) fails to be fulfilled and vanishes if the state described by $%
\mathcal{V}_{m}$ is nonsteerable by Alice's measurements \cite{Kogias1}. For
two-mode Gaussian states, Eq. (\ref{arb}) becomes $\mathcal{G}^{A\rightarrow
B}=\max \left[ 0,\frac{1}{2}\ln \frac{\det \mathcal{V}_{m_{1}}}{4\det 
\mathcal{V}_{m}}\right] $, where $\mathcal{G}^{B\rightarrow A}$ can be
obtained by changing the roles of $A$ and $B$ in Eq. (\ref{arb}) \cite%
{Kogias1}. \newline
Unlike entanglement and Bell non-locality, EPR steering is an asymmetric
aspect of quantum nonlocality,i.e. a quantum state $\varrho _{AB}$ may be
steerable from Alice to Bob, but not vice versa \cite{Kogias1}. Therefore,
we distinguish three cases. The first one corresponding to no-way steering,
where the state is nonsteerable in any direction,i.e. $\mathcal{G}%
^{A\rightarrow B}=\mathcal{G}^{B\rightarrow A}=0$. The second case referring
to two-way steering, where the state is steerable in both directions,i.e. $%
\mathcal{G}^{A\rightarrow B}>0$ and $\mathcal{G}^{B\rightarrow A}>0$.
Finally, the third case in which the state is steerable only in one
direction,i.e. $\mathcal{G}^{A\rightarrow B}>0$ and $\mathcal{G}%
^{B\rightarrow A}=0$ or $\mathcal{G}^{A\rightarrow B}=0$ and $\mathcal{G}%
^{B\rightarrow A}>0$, which corresponds to \textit{one-way steering}. This
last case reflecting the asymmetric nature of quantum correlations is
conjectured to play a decisive role in various communication protocols \cite%
{Kogias2}.\newline
To check how asymmetric can the steerability be in two-mode Gaussian states $%
\varrho _{AB}$, we use the steering asymmetry defined as $\mathcal{G}%
_{AB}^{\Delta }=\left\vert \mathcal{G}^{A\rightarrow B}-\mathcal{G}%
^{B\rightarrow A}\right\vert $ \cite{Kogias1}. It has been proven on the one
hand that $\mathcal{G}_{AB}^{\Delta }$ can never exceed $\ln 2$; it is
maximal when the state is one-way steerable, and it decreases with
increasing steerability in either way, on the other hand, the Gaussian
steering is always upper bounded by the Gaussian R\'{e}nyi-2 entanglement $%
\mathcal{E}_{2}$ with equality on pure states \cite{Kogias1}.

\subsection{Gaussian R\'{e}nyi-2 entanglement}

In quantum information theory, R\'{e}nyi-$\alpha $ entropies are a family of
additive entropies, providing a generalized spectrum of measures of
information in a quantum state $\varrho $ \cite{Renyi}. They are defined as $%
\mathcal{S}_{\alpha }(\varrho )=(1-\alpha )^{-1}\ln \mathrm{Tr}\left(
\varrho ^{\alpha }\right) $, where in the limit $\alpha \rightarrow 1$, $%
\mathcal{S}_{\alpha }(\varrho )$ reduces to the von Neumann entropy $%
\mathcal{S}(\varrho )=-\mathrm{Tr}\left( \varrho \ln \varrho \right) $,
while $\mathcal{S}_{2}(\varrho )=-\ln \mathrm{Tr}\left( \varrho ^{2}\right) $
corresponds to R\'{e}nyi-2 entropy \cite{AGS}. It has been proven that for
Gaussian states, R\'{e}nyi-2 entropy satisfies the strong subadditivity
inequality,i.e. $\mathcal{S}_{2}\left( \mathcal{\varrho }_{AB}\right) +%
\mathcal{S}_{2}\left( \mathcal{\varrho }_{BC}\right) \geqslant \mathcal{S}%
_{2}\left( \mathcal{\varrho }_{ABC}\right) +\mathcal{S}_{2}\left( \mathcal{%
\varrho }_{B}\right) $, therefore it can be used to define valid Gaussian
measures of information and correlation quantities, encompassing
entanglement \cite{AGS}.\newline
For generally mixed two-mode Gaussian states $\mathcal{\varrho }_{AB}$, the R%
\'{e}nyi-2 entanglement measure $\mathcal{E}_{2}\mathcal{(\varrho }_{AB}%
\mathcal{)\equiv E}_{2}$, is not amenable to analytical evaluation and can
only be computed numerically by semidefinite programming \cite{AGS, AL}.
However, for some subclasses of bi-mode Gaussian states including symmetric
states \cite{Giedke and Wolf}, squeezed thermal states (STS) \cite{G(1)} and
GLEMS-Gaussian states of partial minimum uncertainty \cite{AL}, Gaussian R%
\'{e}nyi-2 entanglement (GR2E) can be compactly expressed \cite{AL}. \newline
The covariance matrix $\mathcal{V}_{m}$ (\ref{MCM}) is in the standard form,
where $\mathcal{V}_{m_{12}}=\mathrm{diag}(\upsilon _{12},-\upsilon _{12})$,
which corresponds to \textit{STS} \cite{G(1)}. Thus, the GR2E $\mathcal{E}%
_{2}$ of the two modes $A$ and $B$ with the CM $\mathcal{V}_{m}$ (\ref{MCM})
reads as $\mathcal{E}_{2}=\frac{1}{2}\ln \left[ h(s,d,g)\right] $ with $%
h(s,d,g)=$ $\left[ \frac{(4g+1)s-\sqrt{\left[ (4g-1)^{2}-16d^{2}\right] %
\left[ s^{2}-d^{2}-g\right] }}{4(d^{2}+g)}\right] ^{2}$ if $4|d|+1\leq
4g<4s-1$ and $h(s,d,g)=$ $1$ if $4g\geqslant 4s-1$, where $s=(\upsilon
_{1}+\upsilon _{2})/2,$ $d=(\upsilon _{1}-\upsilon _{2})/2$ and $g=\left(
\upsilon _{1}\upsilon _{2}-\upsilon _{12}^{2}\right) $ \cite{AGS,AL}. 
\newline
The expressions of $\mathcal{G}^{A\rightarrow B}$, $\mathcal{G}%
^{B\rightarrow A}$, $\mathcal{G}_{AB}^{\Delta }$ and $\mathcal{E}_{2}$
involve the covariance matrix elements (\ref{MCM}), which are evaluated as
functions of the squeezing parameter $r$, the $j\mathrm{th}$ optomechanical
cooperativity $\mathcal{C}_{j}$ and the $j\mathrm{th}$ mean thermal photons
number $n_{th,j}$. To observe asymmetric steering, it is necessary to
introduce asymmetry into the system. So, we shall consider the situation
where $n_{th,1}\neq n_{th,2}$ and $\mathcal{C}_{1}\neq \mathcal{C}_{2}$.
Hence, $\mathcal{G}^{A\rightarrow B}$ can not be symmetric by swapping the
roles of $A$ and $B$. Moreover, to have fairly good idea on steering and
entanglement of the two mechanical modes $A$ and $B$, we borrowed realistic
parameters from \cite{Groblacher}. The movable mirrors have a mass $\mu
_{1,2}=145~\mathrm{ng}$, oscillating at frequency $\omega _{m_{1,2}}=2\pi
\times 947~\mathrm{KHz}$ and damped at rate $\gamma _{m_{1,2}}=\gamma
_{m}=2\pi \times 140~\mathrm{Hz}$. The two cavities have length $l_{1,2}=25~%
\mathrm{\ mm}$, decay rate $\kappa _{c_{1,2}}=\kappa _{c}=2\pi \times 215~%
\mathrm{KHz}$, frequency $\omega _{c_{1,2}}=2\pi \times 5.26\times 10^{14}~%
\mathrm{Hz}$ and pumped by lasers of frequency $\omega _{L_{1,2}}=2\pi
\times 2.82\times 10^{14}$ $\mathrm{Hz}$. Notice that the situation $\omega
_{m}\gg \kappa _{c}$, corresponds well to the resolved sideband regime \cite%
{RWA}, which justifies the use of the RWA in section \ref{sec2}.\newline
\begin{figure}[t]
\centerline{\includegraphics[width=0.45\columnwidth,height=5cm]{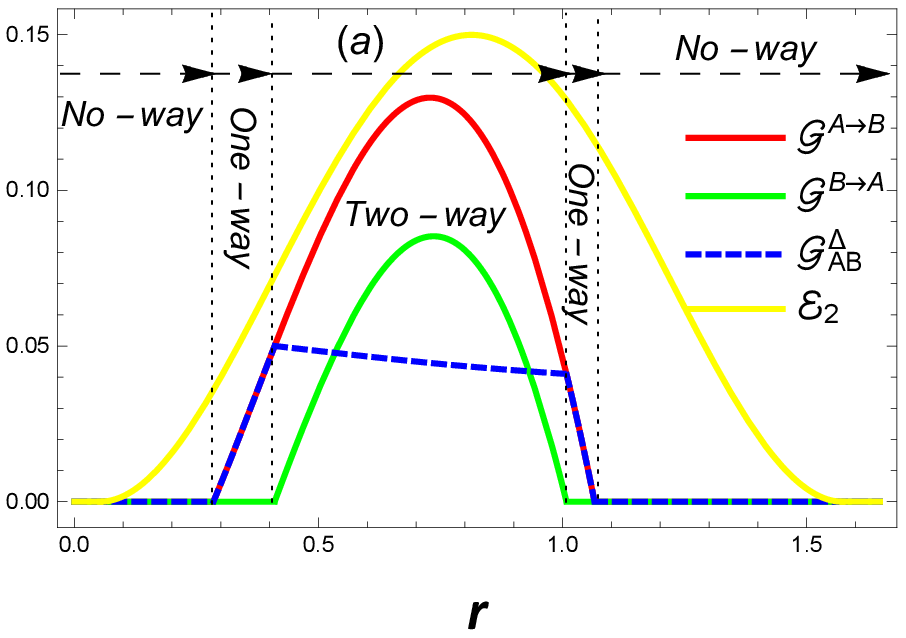}%
\includegraphics[width=0.45\columnwidth,height=5cm]{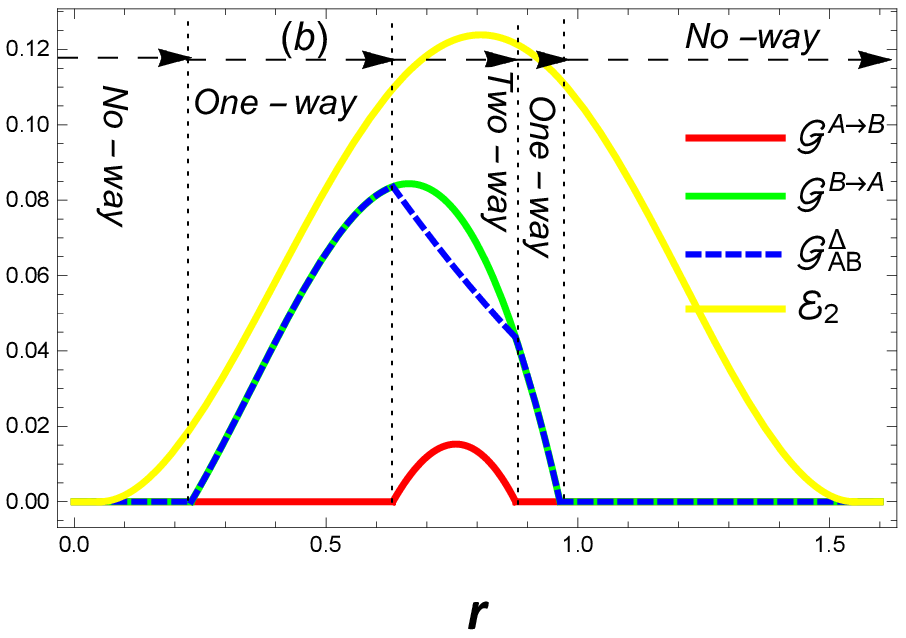}} %
\centerline{\includegraphics[width=0.45\columnwidth,height=5cm]{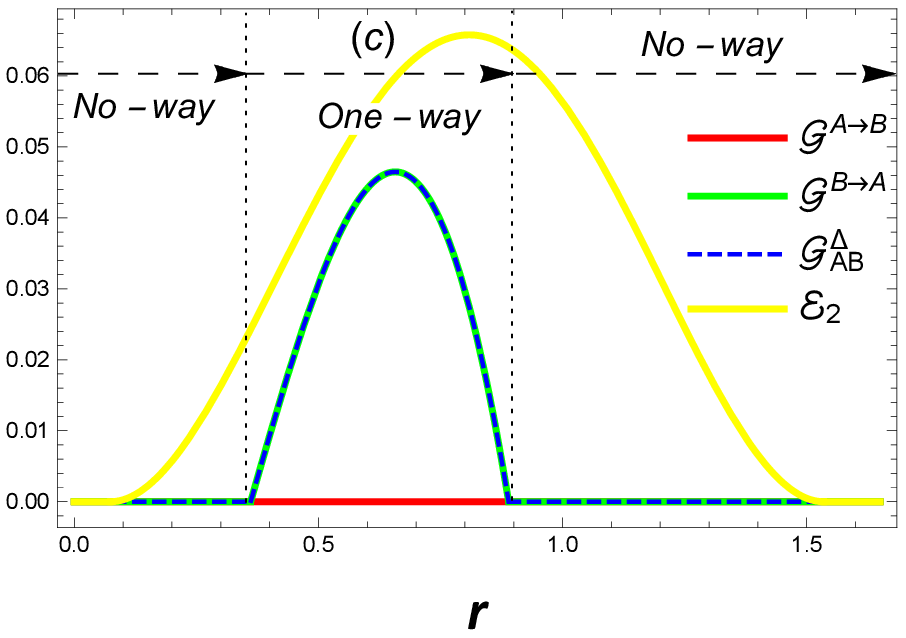}%
\includegraphics[width=0.45\columnwidth,height=5cm]{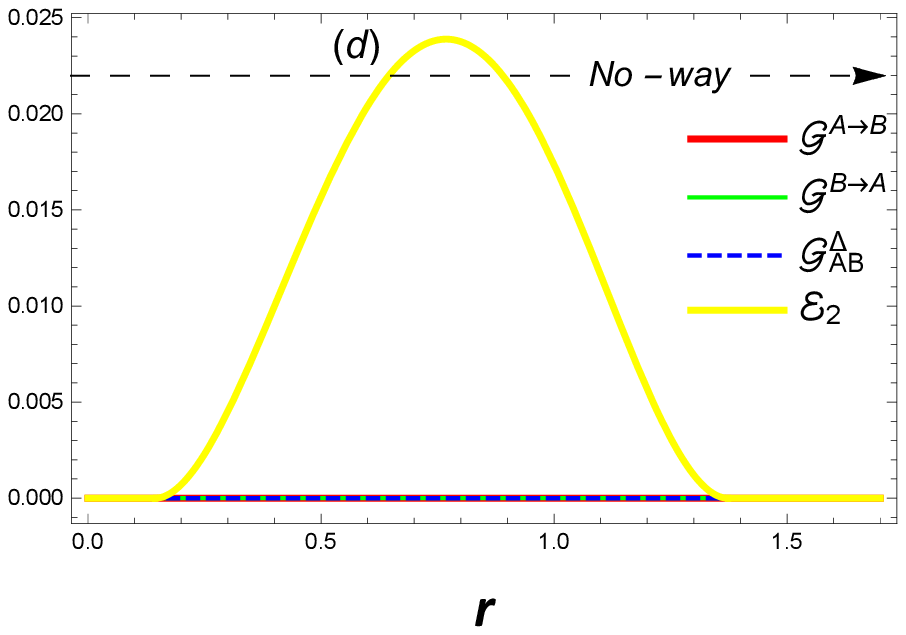}}
\caption{Gaussian steering $\mathcal{G}^{A\rightarrow B}$ (red line), $%
\mathcal{G}^{B\rightarrow A}$ (green line), steering asymmetry $\mathcal{G}%
_{AB}^{\Delta }$ (blue dashed line) and GR2E $\mathcal{E}_{2}$ (yellow line)
of the two modes $A$ and $B$ versus the squeezing $r$ for various thermal
occupations $n_{th,1,2}$. (a) $n_{th,1}=2$, $n_{th,2}=0.5$; (b) $%
n_{th,1}=0.5 $, $n_{th,2}=2$; (c) $n_{th,1}=1$, $n_{th,2}=2$; (d) $%
n_{th,1}=1 $, $n_{th,2}=5$. The optomechanical coupling are $\mathcal{C}%
_{1}=35$ ($\wp _{1}=12~\mathrm{mW}$), $\mathcal{C}_{2}=15$ ($\wp _{2}=5~%
\mathrm{mW}$) \protect\cite{Groblacher}. In panel (c), the states of the two
modes $A$ and $B$ are entangled, however, they are one-way steerable, which
reflects genuinely the asymmetry of quantum correlations.}
\label{Fig.2-}
\end{figure}
In Fig. \ref{Fig.2-}, we consider the squeezing influence on the steering $%
\mathcal{G}^{A\rightarrow B}$, $\mathcal{G}^{B\rightarrow A}$ and
entanglement $\mathcal{E}_{2}$ of the two mechanical modes $A$ and $B$. The
optomechanical cooperativities are fixed as $\mathcal{C}_{1}=35$ and $%
\mathcal{C}_{2}=15$. It should be noted that from Eq. (\ref{OC}), the
condition $\mathcal{C}_{1}\neq \mathcal{C}_{2}$ can be realised for example
by choosing identical parameters for the two cavities, except, the lasers
powers ($\wp _{1}\neq \wp _{2}$).

Fig. \ref{Fig.2-} shows that the steerable states are always entangled,
while, entangled ones are not necessary steerable. Moreover, we remark that
both steering and entanglement undergo the resonance-like behavior under
squeezing effects. This can be well understood knowing that the reduced
state of two-mode squeezed light is a thermal state with a mean photons
number proportional to the squeezing degree $r$ \cite{Paternostro1}. So,
progressive injection of squeezed light increases the photons number in the
two cavities, which leads to strong radiation pressure acting on the
optomechanical coupling. This enhances entanglement and steering of the two
modes $A$ and $B$. On the other hand, steering and entanglement start to
decrease after reaching their maximum. Here the explanation is that, in this
period the photons number becomes important in the two cavities, and
consequently, the thermal noise entering each cavity becomes more
aggressive, bring a quantum correlations degradation. Furthermore, Fig. \ref%
{Fig.2-} reveals that the steering $\mathcal{G}^{A\rightarrow B}$ and $%
\mathcal{G}^{B\rightarrow A}$ are more affected than entanglement $\mathcal{E%
}_{2}$ by thermal noise induced by high squeezing values $r$. In particular,
Figs. \ref{Fig.2-}(a)-\ref{Fig.2-}(b) show that by interchanging the values
of $n_{th,1}$ and $n_{th,2}$, entanglement $\mathcal{E}_{2}$ is not
sensitive to such operation, whereas, the steering $\mathcal{G}%
^{A\rightarrow B}$ and $\mathcal{G}^{B\rightarrow A}$ are strongly affected.
On the other hand, Fig. \ref{Fig.2-}(c) shows an interesting situation in
which the two mechanical modes $A$ and $B$ are entangled for $0.1<r<1.5$,
nonetheless, they are $B\rightarrow A$ \textit{one-way steering}, which
reflects the asymmetry of quantum correlations \cite{WJD}. Such property
could be interpreted as follows: Alice and Bob can perform the same Gaussian
measurements on their shared entangled state, however, obtain contradictory
results. In other words, Bob can convince Alice that their shared state is
entangled, while the converse is not true. This is partly due to the
asymmetry introduced in the system, and partly due to the definition of the
aspect of steering in terms of the EPR paradox \cite{Reid1,Kogias1}.

\begin{figure}[th]
\centerline{\includegraphics[width=0.45\columnwidth,height=5cm]{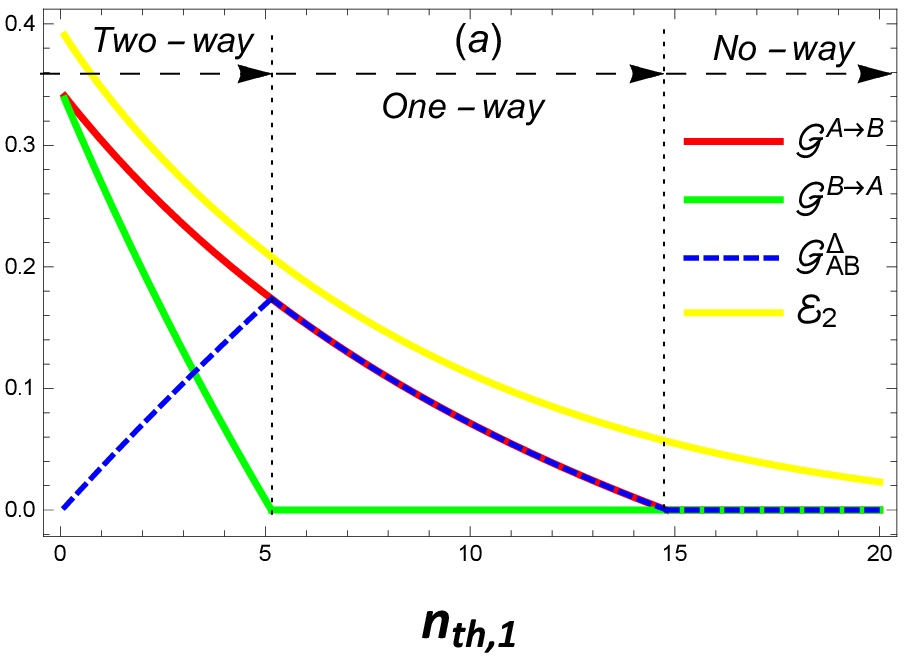}
		\includegraphics[width=0.45\columnwidth,height=5cm]{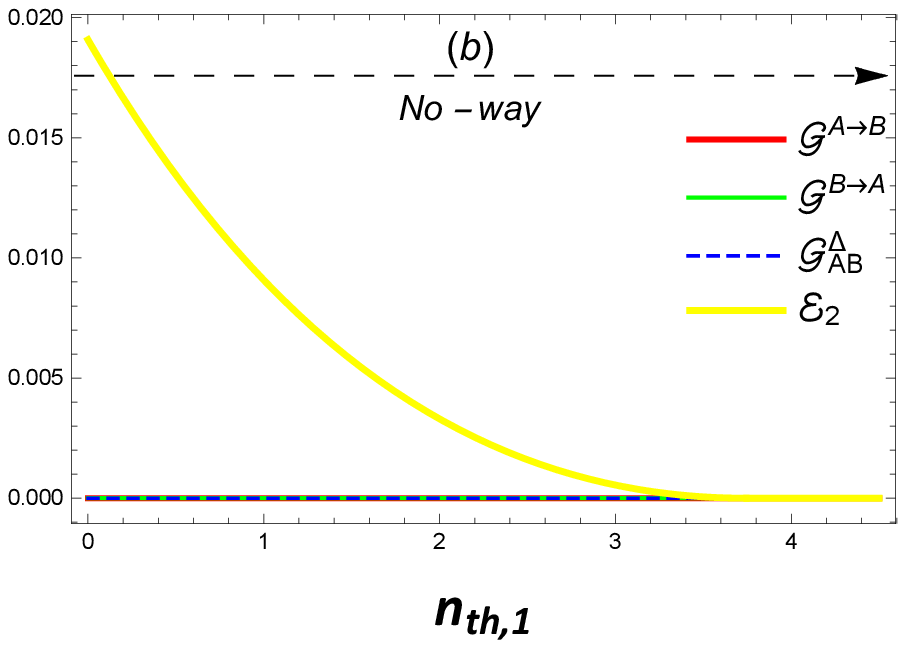}} 
\centerline{\includegraphics[width=0.45\columnwidth,height=5cm]{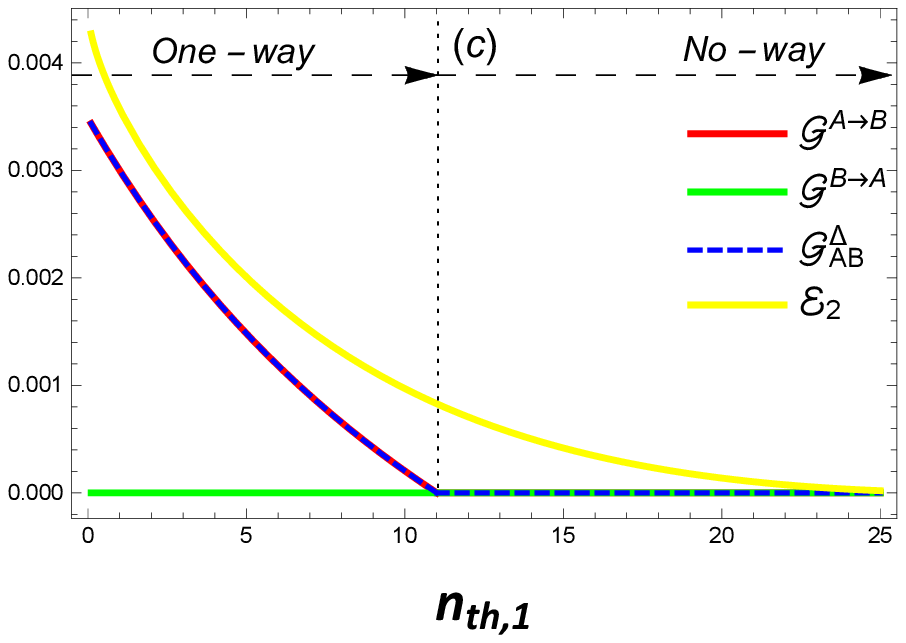}
		\includegraphics[width=0.45\columnwidth,height=5cm]{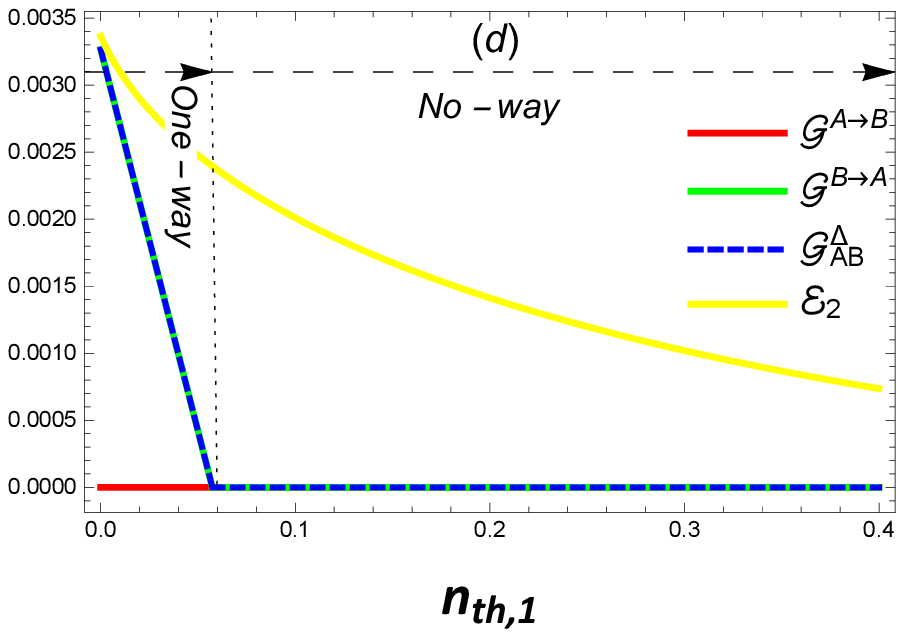}}
\caption{Gaussian steering $\mathcal{G}^{A\rightarrow B}$ (red line), $%
\mathcal{G}^{B\rightarrow A}$ (green line), steering asymmetry $\mathcal{G}%
_{AB}^{\Delta }$ (blue dashed line) and GR2E $\mathcal{E}_{2}$ (yellow line)
of the two modes $A$ and $B$ versus the thermal occupation $n_{th,1}$ for
various values of ${n_{th,2}}$ and squeezing $r$. (a) $r=0.5$, $n_{th,2}=0.1$%
; (b) $r=0.5$, $n_{th,2}=15$; (c) $r=0.05$, $n_{th,2}=0.01$; (d) $r=0.05$, $%
n_{th,2}=1.5$. The optomechanical coupling are $\mathcal{C}_{1}=35$ ($\wp
_{1}=12~\mathrm{mW}$) and $\mathcal{C}_{2}=25$ ($\wp _{2}=8.5~\mathrm{mW}$) 
\protect\cite{Groblacher}. We see that steerable states are entangled but
not necessarily vice versa.}
\label{Fig.3}
\end{figure}
Next, by choosing different values of the squeezing $r$ and thermal
occupations $n_{th,2}$, we consider in Fig. \ref{Fig.3}, the influence of
the thermal occupation $n_{th,1}$ on the steering $\mathcal{G}^{A\rightarrow
B}$, $\mathcal{G}^{B\rightarrow A}$ and entanglement $\mathcal{E}_{2}$. The
optomechanical cooperativities are $\mathcal{C}_{1}=35$ and $\mathcal{C}%
_{2}=25$. Fig. \ref{Fig.3} shows that with increasing $n_{th,1}$, the
steering $\mathcal{G}^{A\rightarrow B}$, $\mathcal{G}^{B\rightarrow A}$
decay strongly than entanglement $\mathcal{E}_{2}$, where they vanish early
than entanglement. Afterwards, fixing the squeezing as $r=0.5$, we remark
from panel \ref{Fig.3}(a) that one-way steering and two-way steering can be
observed with $n_{th,2}=0.1$, while, in panel \ref{Fig.3}(b) with $%
n_{th,2}=15$, the steerability is not authorized in any direction ($\mathcal{%
G}^{A\rightarrow B}=\mathcal{G}^{B\rightarrow A}=0$) although that the two
modes $A$ and $B$ are entangled. This indicates that quantum steering is
more fragile than entanglement against thermal noises. Now, fixing the
squeezing as $r=0.05$, we remark in panel \ref{Fig.3}(c) that for $%
n_{th,2}=0.01$, the two mechanical modes $A$ and $B$ are $A\rightarrow B$
one-way steerable, in contrast, they are $B\rightarrow A$ one-way steerable
for $n_{th,2}=1.5$ in panel \ref{Fig.3}(d). This shows clearly that the
steering of the two modes $A$ and $B$ can be oriented by controlling the
thermal effects. \newline
From Fig. \ref{Fig.3}, it is seen that $\mathcal{G}^{A\rightarrow B}$, $%
\mathcal{G}^{B\rightarrow A}$ and $\mathcal{E}_{2}$ have the same behavior
under the thermal effects,i.e. with increasing of $n_{th,1}$, the steering $%
\mathcal{G}^{A\rightarrow B}$, $\mathcal{G}^{B\rightarrow A}$ decrease with
decreasing of entanglement $\mathcal{E}_{2}$. In fact, such result is not
general, where it has been shown more recently that the decrease of
entanglement does not necessarily mean the decrease of steering \cite%
{Huatang}. Moreover, Fig. \ref{Fig.3} shows that the steerable states are
always entangled, whereas, entangled ones are not necessarily steerable,
which indicates that nonzero degree of entanglement is indispensable for
steering. Further, Figs. \ref{Fig.3}(c)-\ref{Fig.3}(d) show genuine one-way
steering of the two modes $A$ and $B$, indicating that bipartite quantum
correlations are not symmetric in general,i.e. quantum correlations measured
from $A\rightarrow B$ and from $B\rightarrow A$ do not necessarily coincide.
Interesting, Fig. \ref{Fig.3} shows that the steering $\mathcal{G}%
^{A\rightarrow B}$ and $\mathcal{G}^{B\rightarrow A}$ are strongly sensitive
to the thermal noise than entanglement $\mathcal{E}_{2}$, having a tendency
to vanish rapidly with increasing temperature.\newline
Notice that one-way steering observed in Figs. \ref{Fig.2-}-\ref{Fig.3},
answers genuinely the question raised by Wiseman \textit{et al}, i.e. is
then whether there exists bipartite entangled states which are one-way
steerable \cite{WJD}. This most intriguing feature of quantum correlations,
has been expected to play a crucial role in quantum information science,
where it can be employed to guarantee secure quantum communication \cite%
{Saunders}.

Overall, Figs. \ref{Fig.2-}-\ref{Fig.3} show that the steering $\mathcal{G}%
^{A\rightarrow B}$ and $\mathcal{G}^{B\rightarrow A}$ remain upper bounded
by GR2E $\mathcal{E}_{2}$, moreover, the steering asymmetry $\mathcal{G}%
_{AB}^{\Delta }$ is always less than $\ln 2$, it is maximal when the state
is one-way steerable and it decreases with increasing steerability in either
way, which is consistent with \cite{Kogias1}.

\section{Conclusions \label{sec4}}

In an optomechanical system fed by squeezed light and driven in the red
sideband, stationary Gaussian steering of two mechanical modes $A$ and $B$
is studied. In the resolved sideband limit, the steady-state covariance
matrix describing the two considered modes is calculated. We showed that in
the quantum state transfer regime, Gaussian steering can be generated,
while, under influence of thermal effects (squeezing and temperature),
Gaussian one-way steering is occurred genuinely in Figs. \ref{Fig.2-}(c)-\ref%
{Fig.3}[(c)-(d)] and over two phases separated by two-way steering behavior
in Figs. \ref{Fig.2-}[(a)-(b)]. The scenarios illustrated in Figs. \ref%
{Fig.2-}(c)-\ref{Fig.3}[(c)-(d)], where the two modes $A$ and $B$ are
entangled, however, exhibiting genuine one-way steering, reflect the
asymmetric property of quantum correlations,i.e. for a bipartite quantum
state $\varrho _{AB}$, quantum correlations measured from $A\rightarrow B$
and from $B\rightarrow A$ do not necessarily coincide \cite{WJD}. A
comparison study between the steering of the two modes $A$ and $B$ with
their GR2E $\mathcal{E}_{2}$ showed on the one hand, that both steering and
entanglement suffer from a \textit{sudden death}-like phenomenon with early
vanishing of steering in different conditions. On the other hand, Gaussian
steering is found stronger than entanglement, however, remains constantly
upper bounded by Gaussian R\'{e}nyi-2 entanglement, and has a tendency to
decay rapidly to zero under thermal noises. Finally, the steering asymmetry $%
\mathcal{G}_{AB}^{\Delta }$ is found always less than ln2, reaches its
maximum when the two modes $A$ and $B$ are one-way steerable, and it
decreases with increasing steerability in either directions, which is
consistent with \cite{Kogias1}. This work may contribute to the
understanding of the behavior of asymmetric quantum correlations in
dissipative-noisy optomechanical systems, which has immediate applications
in quantum information processing and communication.


\begin{thebibliography}{99}
\bibitem{EPR} A. Einstein, B. Podolsky, and N. Rosen, Phys. Rev. \textbf{47}%
, 777 (1935).

\bibitem{Schrodinger} E. Schr\"{o}dinger, Math. Proc. Cambridge Philos. Soc. 
\textbf{31}, 555 (1935); E. Schr\"{o}dinger, Math. Proc. Cambridge Philos.
Soc. \textbf{32}, 446 (1936).

\bibitem{WJD} H. M.Wiseman, S. J. Jones, and A. C. Doherty, Phys. Rev. Lett. 
\textbf{98}, 140402 (2007).

\bibitem{Bell} J. S. Bell, Physics \textbf{1}, 195 (1964).

\bibitem{The Horodeckis} R. Horodecki, P. Horodecki, M. Horodecki, and K.
Horodecki, Rev. Mod. Phys. \textbf{81}, 865 (2009).

\bibitem{Saunders} D. Saunders, S. Jones, H.Wiseman, and G. Pryde, Nature
Phys. \textbf{6}, 845 (2010); A. J. Bennet, D. A. Evans, D. J. Saunders, C.
Branciard, E. G. Cavalcanti, H. M. Wiseman, and G. J. Pryde, Phys. Rev. X 
\textbf{2}, 031003 (2012).

\bibitem{Reid1} M.D. Reid, Phys. Rev. A \textbf{40}, 913 (1989).

\bibitem{Ou} Z. Y. Ou, S. F. Pereira, H. J. Kimble, and K. C. Peng, Phys.
Rev. Lett. \textbf{68}, 3663 (1992).

\bibitem{Steering inq} E. G. Cavalcanti, S. J. Jones, H. M. Wiseman, and M.
D. Reid, Phys. Rev. A \textbf{80}, 032112 (2009); I. Kogias, P. Skrzypczyk,
D. Cavalcanti, A. Ac\'{\i}n, and G. Adesso, Phys. Rev. Lett. \textbf{115},
210401 (2015); H. Zhu, M. Hayashi, and L. Chen, Phys. Rev. Lett. \textbf{116}%
, 070403 (2016).

\bibitem{Brunner} N.Brunner, D.Cavalcanti, S.Pironio, V.Scarani, and
S.Wehner, Rev. Mod. Phys. \textbf{86}, 419 (2014).

\bibitem{Kogias2} I. Kogias, G. Adesso, J. Opt. Soc. Am. B \textbf{32}, A27
(2015).

\bibitem{Skrzypczyk} P. Skrzypczyk, M. Navascu\'{e}s, and D. Cavalcanti,
Phys. Rev. Lett. \textbf{112}, 180404 (2014).

\bibitem{Watrous} M. Piani and J. Watrous, Phys. Rev. Lett. \textbf{114},
060404 (2015).

\bibitem{Kogias1} I. Kogias, A. R. Lee, S. Ragy, G. Adesso, Phys. Rev. Lett. 
\textbf{114}, 060403 (2015).

\bibitem{Olsen1} M. K. Olsen, Phys. Rev. Lett. \textbf{119}, 160501 (2017).

\bibitem{G1} X.Deng, Y.Xiang, C. Tian, G. Adesso, Q.He, Q.Gong, X.Su, C.Xie
and K. Peng, Phys. Rev. Lett. \textbf{118}, 230501 (2017); J.E. Qars, M.
Daoud and R.A. Laamara. Eur. Phys. J. D \textbf{71}, 122 (2017).

\bibitem{G2} B. Wittmann, S.Ramelow, F.Steinlechner, N.K. Langford, N.
Brunner, H.M. Wiseman, R.Ursin and A. Zeilinger, New J. Phys. \textbf{14},
053030 (2012); K.Sun, X.-J. Ye, J.-S. Xu, X.-Y. Xu, J.-S.Tang, Y.-C.Wu,
J.-L. Chen, C.-F. Li, and G.-C.Guo, Phys. Rev. Lett. \textbf{116}, 160404
(2016); S.Kocsis, M.J. W. Hall, A.J. Bennet, D.J. Saunders and G. J. E.
Pryde, Nat. Commun \textbf{6}, 5886 (2015).

\bibitem{Evans} D. A. Evans and H. M. Wiseman, Phys. Rev. A \textbf{90},
012114 (2014).

\bibitem{Bowles} J. Bowles, T. V\'{e}rtesi, M. T. Quintino, and N. Brunner,
Phys. Rev. Lett. \textbf{112}, 200402 (2014).

\bibitem{oneway} S. L. W. Midgley, A. J. Ferris, and M. K. Olsen, Phys. Rev.
A \textbf{81}, 022101 (2010).

\bibitem{Wollmann} S. Wollmann, N. Walk, A. J. Bennet, H. M. Wiseman, and G.
J. Pryde, Phys. Rev. Lett. \textbf{116}, 160403 (2016).

\bibitem{1sDI} C. Branciard, E. G. Cavalcanti, S. P. Walborn, V. Scarani,
and H. M. Wiseman, Phys. Rev. A \textbf{85}, 010301 (2012); I. Kogias, Y.
Xiang, Q. He, and G. Adesso, Phys. Rev. A \textbf{95}, 012315 (2017).

\bibitem{Reid's} Q. He, L. Rosales-Z\'{a}rate, G. Adesso, and M. D. Reid,
Phys. Rev. Lett. \textbf{115}, 180502 (2015).

\bibitem{Handchen} V. H\"{a}ndchen, T. Eberle, S. Steinlechner, A.
Samblowski, T. Franz, R. F. Werner and R. Schnabel, Nature Photonics \textbf{%
6}, 596 (2012).

\bibitem{Armstrong} S. Armstrong, M. Wang, R.Y. Teh, Q. Gong, Q. He, J.
Janousek, H.-A. Bachor, M.D. Reid and P.K. Lam, Nat. Phys. \textbf{11}, 167
(2015).

\bibitem{QSOS} S. Kiesewetter, Q.Y. He, P.D. Drummond and M.D. Reid, Phys.
Rev. A \textbf{90}, 043805 (2014).

\bibitem{Huatang} H. Tan, W. Deng, Q. Wu, and G. Li. Phys. Rev. A \textbf{95}%
, 053842 (2017).

\bibitem{AGS} G. Adesso, D. Girolami, and A. Serafini, Phys. Rev. Lett. 
\textbf{109}, 190502 (2012).

\bibitem{Meystre} P. Meystre, Ann. Phys. \textbf{525}, 215 (2013).

\bibitem{Aspelmeyer} M. Aspelmeyer, T.J. Kippenberg and F. Marquardt, Rev.
Mod. Phys. \textbf{86}, 1391 (2014).

\bibitem{Cooling} J. D. Teufel, T. Donner, D. Li, J. W. Harlow, M. S.
Allman, K. Cicak, A. J. Sirois, J. D. Whittaker, K. W. Lehnert and R. W.
Simmonds, Nature (London) \textbf{475}, 359 (2011).

\bibitem{Agarwal} G.S. Agarwal and S. Huang, Phys. Rev. A \textbf{81},
041803(R) (2010).

\bibitem{Liao} X. -Y. L\"{u} , J. -Q. Liao , L. Tian, F. Nori, Phys. Rev. A 
\textbf{91}, 013834(7) (2015).

\bibitem{Marshall} W. Marshall, C. Simon, R. Penrose, D. Bouwmeester, Phys.
Rev. Lett. \textbf{91}, 130401 (2003).

\bibitem{Korppi} C. F. Ockeloen-Korppi, et al., Phys. Rev. Lett. \textbf{117}%
, 140401 (2016).

\bibitem{entanglement} T. A. Palomaki, J. D. Teufel, R. W. Simmonds, and K.
W. Lehnert, Science \textbf{342}, 710 (2013); J. El Qars, M. Daoud, Ahl
Laamara, Int. J. Quant. Inform. \textbf{13}, 1550041 (2015); J. El Qars, M.
Daoud, R. Ahl Laamara, Int. J. Mod. Phys. B \textbf{30}, 1650134 (2016); J.
El Qars, M. Daoud and R. Ahl Laamaraa, J. Mod. Opt \textbf{65}, 1584 (2018).

\bibitem{Law} C. K. Law, Phys. Rev. A \textbf{51}, 2537 (1995).

\bibitem{Paternostro} M. Paternostro, L. Mazzola, and J. Li, J. Phys. B: At.
Mol. Opt. Phys. \textbf{45}, 154010 (2012).

\bibitem{Genes} C. Genes, A. Mari, D. Vitali, P. Tombesi, Adv. At. Mol. Opt.
Phys. \textbf{57}, 33 (2009).

\bibitem{Benguria} R. Benguria, and M. Kac, Phys. Rev. Lett, \textbf{46}, 1
(1981).

\bibitem{Parkin and Kimble} A. S. Parkins and H. J. Kimble, J. Opt. B:
Quantum Semiclass. Opt. \textbf{1} 496 (1999).

\bibitem{RWA} Y.-D. Wang, S. Chesi, A. A. Clerk, Phys. Rev. A \textbf{91},
013807 (2015).

\bibitem{Parks} P. C. Parks and V. Hahn, Stability Theory. New York:
Prentice Hall, 1993.

\bibitem{OC} T. P. Purdy, P.-L. Yu, R. W. Peterson, N. S. Kampel, and C. A.
Regal, Phys. Rev. X \textbf{3}, 031012 (2013).

\bibitem{G(1)} G. Adesso and A. Datta, Phys. Rev. Lett. \textbf{105}, 030501
(2010).

\bibitem{AL} G. Adesso and F. Illuminati, Phys. Rev. A \textbf{72}, 032334
(2005).

\bibitem{Lami1} L. Lami, C. Hirche, G. Adesso, and A. Winter, Phys. Rev.
Lett. \textbf{117}, 220502 (2016).

\bibitem{Renyi} A. R\'{e}nyi "On measures of information and entropy",
Proceedings of the 4th Berkeley Symposium on Mathematics, Statistics and
Probability, p 547; 1960.

\bibitem{Giedke and Wolf} G. Giedke, M.M. Wolf, O. Kr\"{u}ger, R.F.Werner
and J.I. Cirac, Phys. Rev. Lett. \textbf{91}, 107901 (2003).

\bibitem{Groblacher} S. Gr\"{o}blacher, K. Hammerer, M.R. Vanner and M.
Aspelmeyer, Nature(London) \textbf{460}, 724 (2009).

\bibitem{Paternostro1} L. Mazzola, M. Paternostro, Phys. Rev. A \textbf{83},
062335 (2011).
\end{thebibliography}
\end{document}